\documentclass[aps,prl,twocolumn,superscriptaddress]{revtex4-1}
\usepackage{graphicx}
\usepackage{epstopdf}
\usepackage{dcolumn}
\usepackage{bm}
\usepackage{subfigure}
\usepackage{amsmath}
\usepackage{epsfig}
\usepackage{verbatim}
\usepackage{float}
\usepackage{url}
\usepackage{color} 

\begin{document}
\title{Optimal vaccination program for two infectious diseases with cross immunity}
\author{Yang Ye}
\author{Qingpeng Zhang}
\email{qingpeng.zhang@cityu.edu.hk}
\affiliation{School of Data Science, City University of Hong Kong, Hong Kong SAR, China}

\author{Zhidong Cao}
\affiliation{The State Key Laboratory of Management and Control for Complex Systems, Institute of Automation, Chinese Academy of Sciences, Beijing, China}
\affiliation{School of Artificial Intelligence, University of Chinese Academy of Sciences, Beijing, China}
\affiliation{Shenzhen Artificial Intelligence and Data Science Institute (Longhua), Shenzhen, China}

\author{Daniel Dajun Zeng}
\affiliation{The State Key Laboratory of Management and Control for Complex Systems, Institute of Automation, Chinese Academy of Sciences, Beijing, China}
\affiliation{School of Artificial Intelligence, University of Chinese Academy of Sciences, Beijing, China}
\affiliation{Shenzhen Artificial Intelligence and Data Science Institute (Longhua), Shenzhen, China}

\begin{abstract}
There are often multiple diseases with cross immunity competing for vaccination resources. Here we investigate the optimal vaccination program in a two-layer Susceptible-Infected-Removed (SIR) model, where two diseases with cross immunity spread in the same population, and vaccines for both diseases are available. We identify three scenarios of the optimal vaccination program, which prevents the outbreaks of both diseases at the minimum cost. We analytically derive a criterion to specify the optimal program based on the costs for different vaccines.

\end{abstract}

\maketitle
Vaccination is an effective method of preventing epidemics like COVID-19, diphtheria, influenza, and measles \cite{andre2008vaccination,bonanni1999demographic,le2020covid}. Mathematical models have been commonly used to examine the effect of vaccination on epidemic control and prevention \cite{wang2016statistical,PhysRevLett.101.078101,PhysRevE.101.062306}. Vaccines help the human body's natural defense systems to develop antibodies to pathogens. Generally, antibodies to one pathogen do not provide protection to another pathogen. However, there is increasing evidence of cross immunity between diseases, where exposure to one pathogen may help protect against other pathogens. \cite{wu2020interference,schultz2015viral,BALMER2011868,laurie2015interval,laurie2018evidence}. Cross immunity is a common outcome when some diseases caused by related pathogens spread in the same population. There is rich previous epidemiological research on the interactions pathogens with cross immunity from the perspectives of age, gender, and population structure \cite{andreasen1997dynamics,castillo1989epidemiological,castillo1996competitive,li2003coexistence,webb2013role,leventhal2015evolution,PhysRevE.89.062817,PhysRevE.93.042303}. Some physical literature provides analytic calculations of the expected final epidemic sizes of multiple diseases coexisting in the same population \cite{PhysRevE.84.036106, PhysRevLett.95.108701}. 
It is needed to take into account the effect of cross immunity while developing disease prevention strategies. Some studies gave simulation and optimization frameworks to identify the effective resource allocation to control several competitive diseases in the susceptible-infected-susceptible (SIS) model \cite{watkins2016optimal,7040362}. However, the identification of the optimal vaccination programs to control competitive diseases with cross immunity in the susceptible-infected-recovered (SIR) model is under-researched and critically-needed, particularly when the world is expecting to face the challenge of allocating COVID-19 vaccines in the coming years. 

In this letter, we assume that there are two vaccines for two SIR type diseases with varying cost. A vaccination program specifies the fractions of individuals that should be vaccinated for each vaccine. We define that the optimal vaccination program can control the outbreak of both diseases at the minimum cost. The microscopic Markov chain approach (MMCA) is adopted to derive the conditions guaranteeing that both diseases can be contained in the population. An optimization framework is proposed to analytically derive the optimal vaccination program. We identify three scenarios of the optimal vaccination program, and derive a criterion to specify the optimal program based on the costs for different vaccines.

Here, we consider a two-layer network with the same topological structure, which represents the transmission channels for both diseases, $i=1,2$, which propagate through different layers. Nodes represent individuals, which are susceptible ($S_i$), infected ($I_i$), or recovered ($R_i$) for disease $i$. A susceptible node $S_i$ can be infected by an infected node $I_i$ with an infection rate $\beta_i$. An infected node $I_i$ has a transition rate $\gamma_i$ to become recovered and immune to disease $i$. The susceptibility to disease $j$ ($j\neq i$) of individuals recovered from disease $i$ is reduced by a factor $\alpha_j\in(0,1)$ (i.e., the actual infection rate for disease $j$ becomes $\alpha_j\beta_j$). $1-\alpha_j$ quantifies the level of cross immunity. In the extreme case, $\alpha_j=0$ corresponds to complete cross immunity, and $\alpha_j=1$ corresponds to a simpler problem without cross immunity. An individual can be simultaneously infected by both diseases. Before the occurrence of the first node infected with disease $i$, a fraction of $v_i\in[0,1]$ nodes are provided with vaccines for disease $i$. We assume that vaccine-induced immunity is equivalent to the natural immunity obtained from actual infection. Thus, vaccines for disease $i$ directly transfer the node state from $S_i$ to $R_i$ without causing illness. Nodes vaccinated with vaccines for disease $i$ are also immune to disease $i$ and have a susceptibility to disease $j$ ($j\neq i$) reduced by a factor $\alpha_j$. After vaccination, a small fraction of unvaccinated nodes become initially infected for both diseases. 

\begin{figure}[htbp]
\includegraphics[width=1.0\columnwidth]{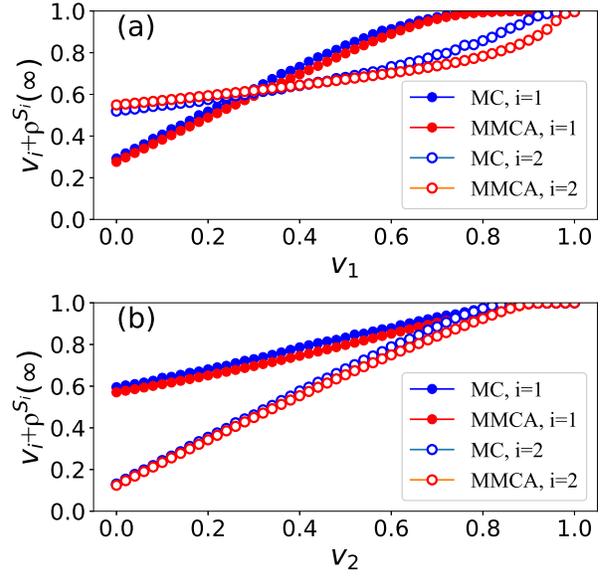}
\caption{\label{MMCA_MC}Comparison of the fraction of uninfected nodes at the stationary state [$v_i+\rho^{S_i}(\infty)$, $i=1,2$] using MC simulations and MMCA. We consider a network of 2000 nodes with a Poisson degree distribution, where the mean degree is 20. Results are obtained by averaging 1000 MC simulations. (a) $v_2=0.5$ while changing the values of $v_1$ and (b) $v_1=0.5$ while changing the values of $v_2$. Parameter values $\beta_1=0.4$, $\gamma_1=0.8$, $\alpha_1=0.1$, $\beta_2=0.5$, $\gamma_2=0.6$, and $\alpha_2=0.1$.}
\end{figure}

According to this scheme, every node can be in nine different states at each time: $S_1S_2$, $S_1I_2$, $S_1R_2$, $I_1S_2$, $I_1I_2$, $I_1R_2$, $R_1S_2$, $R_1I_2$, and $R_1R_2$. Note that the character order does not affect the state (e.g., $S_1S_2$ is equivalent to $S_2S_1$). Every node $n$ has a certain probability of being in one of the nine states at time $t$ denoted by $p^{S_1S_2}_n(t)$, $p^{S_1I_2}_n(t)$, $p^{S_1R_2}_n(t)$, $p^{I_1S_2}_n(t)$, $p^{I_1I_2}_n(t)$, $p^{I_1R_2}_n(t)$, $p^{R_1S_2}_n(t)$, $p^{R_1I_2}_n(t)$, and $p^{R_1R_2}_n(t)$. $p^{S_1S_2}_n(t)+p^{S_1I_2}_n(t)+p^{S_1R_2}_n(t)+p^{I_1S_2}_n(t)+p^{I_1I_2}_n(t)+p^{I_1R_2}_n(t)+p^{R_1S_2}_n(t)+p^{R_1I_2}_n(t)+p^{R_1R_2}_n(t)=1$. For disease $i$ at time $t$, node $n$ has a certain probability of being in one of three states $S_i$, $I_i$, or $R_i$, denoted by 
\begin{equation}
    \begin{aligned}
     p_n^{S_i}(t) &= p_n^{S_iS_j}(t) + p_n^{S_iI_j}(t)+p_n^{S_iR_j}(t),\\
  p_n^{I_i}(t) &= p_n^{I_iS_j}(t) + p_n^{I_iI_j}(t)+p_n^{I_iR_j}(t),\\
    p_n^{R_i}(t) &= p_n^{R_iS_j}(t) + p_n^{R_iI_j}(t)+p_n^{R_iR_j}(t),
    \end{aligned}
\label{p_n_single}
\end{equation}
where $i\neq j$. $p_n^{R_i}(0)=v_i$ due to the effect of vaccination. If node $n$ is recovered from disease $j$, the probability of not being infected for disease $i$ ($i\neq j$) is 
\begin{equation}
    q_{n,i}^P(t) = \prod\limits_{m}(1-a_{mn}p_m^{I_i}(t)\alpha_i\beta_i),
\label{qp}
\end{equation}
where $a_{mn}$ is the adjacency matrix on both layers. If node $n$ is not recovered from disease $j$, the probability is 
\begin{equation}
    q_{n,i}^U(t) = \prod\limits_{m}(1-a_{mn}p_m^{I_i}(t)\beta_i).
\label{qu}
\end{equation}
We develop the microscopic Markov chain approach to analyze the state transitions for each node $n$ as
\begin{equation}
\begin{split}
p_n^{S_1S_2}(t+1) &= q_{n,1}^U(t)q_{n,2}^U(t)p_n^{S_1S_2}(t),\\
p_n^{S_1I_2}(t+1) &= q_{n,1}^U(t)[1-q_{n,2}^U(t)]p_n^{S_1S_2}(t)\\
&\quad+q_{n,1}^U(t)(1-\gamma_2)p_n^{S_1I_2}(t),\\
p_n^{S_1R_2}(t+1) &= q_{n,1}^U(t)\gamma_2p_n^{S_1I_2}(t)+q_{n,1}^P(t)p_n^{S_1R_2}(t),\\
p_n^{I_1S_2}(t+1) &= [1-q_{n,1}^U(t)]q_{n,2}^U(t)p_n^{S_1S_2}(t)\\
&\quad+(1-\gamma_1)q_{n,2}^U(t)p_n^{I_1S_2}(t),\\
p_n^{I_1I_2}(t+1) &= [1-q_{n,1}^U(t)][1-q_{n,2}^U(t)]p_n^{S_1S_2}(t)\\
&\quad+[1-q_{n,1}^U(t)](1-\gamma_2)p_n^{S_1I_2}(t)\\
&\quad+(1-\gamma_1)[1-q_{n,2}^U(t)]p_n^{I_1S_2}(t)\\
&\quad+(1-\gamma_1)(1-\gamma_2)p_n^{I_1I_2}(t),\\
p_n^{I_1R_2}(t+1) &= [1-q_{n,1}^U(t)]\gamma_2p_n^{S_1I_2}(t)\\
&\quad+[1-q_{n,1}^P(t)]p_n^{S_1R_2}(t)\\
&\quad+(1-\gamma_1)\gamma_2p_n^{I_1I_2}(t)\\
&\quad+(1-\gamma_1)p_n^{I_1R_2}(t),\\
p_n^{R_1S_2}(t+1) &= \gamma_1q_{n,2}^U(t)p_n^{I_1S_2}(t)+q_{n,2}^P(t)p_n^{R_1S_2}(t),\\
p_n^{R_1I_2}(t+1) &= \gamma_1[1-q_{n,2}^U(t)]p_n^{I_1S_2}(t)\\
&\quad+\gamma_1(1-\gamma_2)p_n^{I_1I_2}(t)\\
&\quad+(1-q_{n,2}^P(t))p_n^{R_1S_2}(t)\\
&\quad+(1-\gamma_2)p_n^{R_1I_2}(t),\\
p_n^{R_1R_2}(t+1) &= \gamma_1\gamma_2p_n^{I_1I_2}(t)+\gamma_1p_n^{I_1R_2}(t)\\
&\quad+\gamma_2p_n^{R_1I_2}(t)+p_n^{R_1R_2}(t).
\end{split}
\label{MMCA}
\end{equation}
 The fractions of susceptible, infected, and recovered nodes at time $t$ for disease $i$ is denoted by $\rho^{S_i}(t)$, $\rho^{I_i}(t)$, and $\rho^{R_i}(t)$, respectively. 
\begin{equation}
\begin{aligned}
\rho^{S_i}(t)=\frac{1}{N}\sum\limits_np_n^{S_i}(t),\\
\rho^{I_i}(t)=\frac{1}{N}\sum\limits_np_n^{I_i}(t),\\
\rho^{R_i}(t)=\frac{1}{N}\sum\limits_np_n^{R_i}(t).
\end{aligned}
\end{equation}
Here $N$ is the number of nodes on the network. The fraction of uninfected nodes at the stationary state (when $t\rightarrow\infty$) is $\rho^{R_i}(0)+\rho^{S_i}(\infty)=v_i+\rho^{S_i}(\infty)$ for disease $i$. We perform extensive Monte Carlo simulations (MC) to validate
the MMCA results obtained by Eqs.~(\ref{MMCA}). We adopt the synchronous updating method in MC simulations. All nodes update their
states simultaneously in each time step, which is set as 1. We consider a network
of 2000 nodes with a Poisson degree distribution, where the mean degree is 20. The initial fraction of infected nodes is $\frac{1}{N}=0.0005$ for diseases $i$ if $v_i\neq 1$. Each point in Fig.~\ref{MMCA_MC} is obtained
by averaging 1000 MC simulations. The average $R^2$ is 0.96, indicating a high agreement between MMCA and MC simulation results.

Next, we explore the conditions preventing the outbreaks of both diseases. Specifically, the outbreak of disease $i$ can be prevented when
\begin{equation}
    \left.\dfrac{d\rho^{I_i}(t)}{dt}\right|_{t=0} < 0,
\end{equation}
\begin{equation}
\begin{split}
    \left.\dfrac{d\rho^{I_i}(t)}{dt}\right|_{t=0}=\rho^{I_i}(1)-\rho^{I_i}(0)=\frac{\sum\limits_n p_n^{I_i}(1)-p_n^{I_i}(0)}{N}.
\end{split}
\end{equation}
According to Eq.~(\ref{p_n_single}) and Eq.~(\ref{MMCA}),
\begin{equation}
\begin{split}
    p_n^{I_i}(1)&=p_n^{I_iS_j}(1)+p_n^{I_iI_j}(1)+p_n^{I_iR_j}(1)\\
    &=[p_n^{S_iS_j}(0)+p_n^{S_iI_j}(0)][1-q_{n,i}^U(0)]\\
    &\quad+p_n^{S_iR_j}(0)[1-q_{n,i}^P(0)]+(1-\gamma_i)p_n^{I_i}(0).
\end{split}
\end{equation}
Thus,
\begin{equation}
\begin{split}
    \left.\dfrac{d\rho^{I_i}(t)}{dt}\right|_{t=0}&=\frac{1}{N}\sum\limits_n \{[1-q_{n,i}^U(0)][p_n^{S_iS_j}(0)\\
    &\quad+p_n^{S_iI_j}(0))]+[1-q_{n,i}^P(0)]p_n^{S_iR_j}(0)\\
    &\quad-\gamma_ip_n^{I_i}(0)\}.
\end{split}
\end{equation}
If $0\leq v_i<1$, $p_n^{I_i}(0)=\frac{1}{N}$, $p_n^{S_i}(0)=1-v_i-\frac{1}{N}$, and $p_n^{R_i}(0)=0$. Thus, $p_n^{S_iS_j}=(1-v_i-\frac{1}{N})(1-v_j-\frac{1}{N})$, $p_n^{S_iI_j}=(1-v_i-\frac{1}{N})\frac{1}{N}$, and $p_n^{S_iR_j}=(1-v_i-\frac{1}{N})v_j$. If $N$ is large enough, $p_n^{I_i}(0)=\frac{1}{N}\rightarrow0$,  consequently, from Eq.~(\ref{qp}) and Eq.~(\ref{qu}), we obtain
\begin{equation}
    \begin{split}
    q_{n,i}^P(0)&\approx1-\frac{\alpha_i\beta_i}{N}\sum\limits_ma_{mn}=1-\frac{\alpha_i\beta_i}{N}k_n,\\
        q_{n,i}^U(0) &\approx1-\frac{\beta_i}{N}\sum\limits_ma_{mn}=1-\frac{\beta_i}{N}k_n,
    \end{split}
\end{equation}
where $k_n$ is the degree of node $n$. Thus, 
\begin{equation}
\begin{split}
\left.\dfrac{d\rho^{I_i}(t)}{dt}\right|_{t=0}&\approx \frac{\overline{k}\beta_i}{N}(1-v_i)[(1-v_j)+\alpha_iv_j]\\
&\quad-\frac{\gamma_i}{N},
\end{split}
\label{di}
\end{equation}
where $\overline{k}$ is the mean degree of the network. To ensure the failure of outbreaks for both diseases, the right-hand side of Eq.~(\ref{di}) should be less or equal to 0, because it is slightly larger than the left-hand side. Therefore, the following conditions should be met
\begin{equation}
\begin{aligned}
    R_{v,1}&=R_{0,1}(1-v_1)[1-(1-\alpha_1)v_2]\leq1,\\
    R_{v,2}&=R_{0,2}(1-v_2)[1-(1-\alpha_2)v_1]\leq1,
\end{aligned}
\label{rv conditions}
\end{equation} 
where $R_{0,1}=\frac{\beta_1\overline{k}}{\gamma_1}$ and $R_{0,2}=\frac{\beta_2\overline{k}}{\gamma_2}$. $R_{0,1}$ and $R_{0,2}$ are the basic reproduction numbers for disease 1 and disease 2, respectively. If $v_i=1$, $\rho_n^{I_i}(0)=\rho_n^{S_i}(0)=\rho_n^{R_i}(0)=0$, disease $i$ can never spread out and Eqs.~(\ref{rv conditions}) still hold. Here $R_{v,i}$ is the expected number of infections caused by an individual infected with disease $i$ when the fractions of vaccinated individuals are $v_1$ and $v_2$ at the beginning. Note that this corresponds to $R_{v,i}=R_{0,i}$ for $v_1=v_2=0$. According to Eqs.~(\ref{rv conditions}), in order to prevent the outbreaks of both diseases, the following conditions should be met 
\begin{equation}
\begin{aligned}
    v_1 & \geq 1- \frac{1}{R_{0,1}[1-(1-\alpha_1)v_2]},\\
    v_2 & \geq 1- \frac{1}{R_{0,2}[1-(1-\alpha_2)v_1]},
\end{aligned}
\label{conditions}
\end{equation}
From Eqs.~(\ref{conditions}), we observe that fewer vaccines for disease $i$ are needed to prevent the outbreak of disease $i$ if more nodes are vaccinated for disease $j$ ($j\neq i$). When $v_j=1$ (all nodes are vaccinated for disease $j$), 
$v_i \geq 1-\frac{1}{R_{0,i}\alpha_i}$, indicating that the minimum fraction of nodes that should be vaccinated for disease $i$ to prevent the outbreak of disease $i$ is $1-\frac{1}{R_{0,i}\alpha_i}$. We define this value as
\begin{equation}
    \tilde{v}_i=1-\frac{1}{R_{0,i}\alpha_i}.
\label{critical fraction}
\end{equation}
In the following, we consider the common scenario when $R_{0,i}>1$ and $\tilde{v}_i\geq0$. The results can be easily extended to other scenarios.

\begin{figure}
\includegraphics[width=1.0\columnwidth]{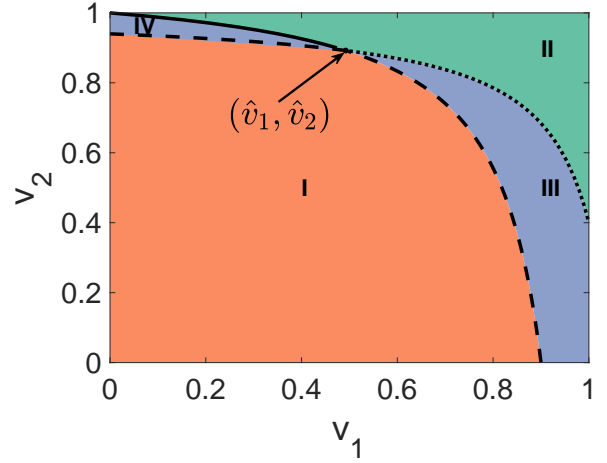}
\caption{\label{outbreak}Full $v_2-v_1$ phase diagram for the same multiplex in Fig.~\ref{MMCA_MC}. Region \uppercase\expandafter{\romannumeral1}: both
diseases can outbreak; Region \uppercase\expandafter{\romannumeral2}: both
diseases cannot outbreak; Region \uppercase\expandafter{\romannumeral3}: disease 1 cannot outbreak, while disease 2 can outbreak; Region \uppercase\expandafter{\romannumeral4}:  disease 2 cannot outbreak, while disease 1 can outbreak. The transition lines are computed from Eqs.~(\ref{conditions}). The intersection $(\hat{v}_1, \hat{v}_2)$ is computed from Eqs.~(\ref{critical point}). Parameters are as in Fig.~\ref{MMCA_MC}.}
\end{figure}

Now we plot the full $v_2-v_1$ phase diagram of the outbreak conditions for both diseases in Fig.~\ref{outbreak}. We adopt the same two-layer network setting as Fig.~\ref{MMCA_MC}. As illustrated in Fig.~\ref{outbreak}, only points of ($v_1, v_2$) in region \uppercase\expandafter{\romannumeral2} can prevent the outbreaks of both diseases. The intersection $(\hat{v}_1, \hat{v}_2)$ satisfies that
\begin{equation}
\begin{aligned}
    \hat{v}_1 & = 1- \frac{1}{R_{0,1}[1-(1-\alpha_1)\hat{v}_2]},\\
    \hat{v}_2 & = 1- \frac{1}{R_{0,2}[1-(1-\alpha_2)\hat{v}_1]}.
\end{aligned}
\end{equation}
Denote $\Delta =[\frac{1-\alpha_2}{R_{0,1}}-\frac{1-\alpha_1}{R_{0,2}}]^2+\alpha_1^2\alpha_2^2+2\alpha_1\alpha_2[\frac{1-\alpha_2}{R_{0,1}}+\frac{1-\alpha_1}{R_{0,2}}]$,
\begin{equation}
\begin{split}
    \hat{v}_1 & = \frac{1}{2}+\frac{\alpha_1-\sqrt{\Delta}}{2\alpha_1(1-\alpha_2)}+\frac{1-\alpha_1}{2R_{0,2}\alpha_1(1-\alpha_2)}\\
    &\quad-\frac{1}{2R_{0,1}\alpha_1}, \\
    \hat{v}_2 & = \frac{1}{2}+\frac{\alpha_2-\sqrt{\Delta}}{2\alpha_2(1-\alpha_1)}+\frac{1-\alpha_2}{2R_{0,1}\alpha_2(1-\alpha_1)}\\
    &\quad-\frac{1}{2R_{0,2}\alpha_2}.
\end{split}
\label{critical point}
\end{equation}

Next, we aim to identify the minimum vaccination cost that can prevent the outbreaks of both diseases. Denote $C_1$ and $C_2$ as the costs of vaccination per capita for disease 1 and disease 2, respectively. $C_1$ and $C_2$ are positive. The total vaccination cost is $V=C_1v_1N+C_2v_2N$. Then, we formulate the following optimization problem
\begin{equation}
\begin{aligned}
& {\text{minimize}} & & V \\
& \text{subject to} & & v_1 \geq 1- \frac{1}{R_{0,1}[1-(1-\alpha_1)v_2]},\\
&&&v_2 \geq 1- \frac{1}{R_{0,2}[1-(1-\alpha_2)v_1]},\\
&&&0\leq v_1 \leq 1,\\
&&&0\leq v_2 \leq 1.
\end{aligned}
\end{equation}
Since the slopes for $v_1$ and $v_2$ are both positive, the optimal vaccination program ($v_1^*$, $v_2^*$) is on the boundary of region \uppercase\expandafter{\romannumeral2} in Fig.~\ref{outbreak}. 

On the solid line (the boundary between region \uppercase\expandafter{\romannumeral2} and region \uppercase\expandafter{\romannumeral4}),  
\begin{equation*}
    v_1=1- \frac{1}{R_{0,1}[1-(1-\alpha_1)v_2]}
\end{equation*}
and $v_2\in[\hat{v}_2, 1]$, thus, \begin{equation}
    V = C_1\{1-\frac{1}{R_{0,1}[1-(1-\alpha_1)v_2]}\}N + C_2v_2N,
\end{equation}
\begin{equation}
    \frac{\mathrm{d}^2V}{\mathrm{d}v_2^2}=-\frac{2C_1(1-\alpha_1)^2N}{R_{0,1}[1-(1-\alpha_1)v_2]^3}<0,
\end{equation}
Therefore, $V$ is a concave function of $v_2$. 
The optimal vaccination program ($v_1^*$, $v_2^*$) is either $(\hat{v}_1, \hat{v}_2)$ or $(\tilde{v}_1, 1)$. Similarly, on the dotted line (the boundary between region \uppercase\expandafter{\romannumeral2} and region \uppercase\expandafter{\romannumeral3}), the optimal vaccination program ($v_1^*$, $v_2^*$) is either $(\hat{v}_1, \hat{v}_2)$ or $(1, \tilde{v}_2)$. Overall, there are three potential scenarios of the optimal vaccination program: (1, $\tilde{v}_2$), ($\tilde{v}_1$, 1), and ($\hat{v}_1, \hat{v}_2$). 

Comparing the values of $V$ on these points, we find that the value of $(v_1^*, v_2^*)$ is determined by the following four nonnegative parameters:
\begin{align}
   L_1   &= 1-\hat{v}_1, &L_2&=1-\hat{v}_2,\\
  L_3&=\hat{v}_1 - \tilde{v}_1, &L_4&=\hat{v}_2 - \tilde{v}_2.
\label{l}
\end{align}
If $L_1L_2\geq L_3L_4$, we have
\begin{equation*}
    \frac{L_4}{L_1}\leq \frac{L_2+L_4}{L_1+L_3}\leq\frac{L_2}{L_3},
\end{equation*}
thus,
\begin{equation}
(v_1^*, v_2^*)=\left\{
\begin{aligned}
(1, \tilde{v}_2)\quad&\frac{C_1}{C_2}<\frac{L_4}{L_1},\\
(\hat{v}_1, \hat{v}_2)\quad&\frac{L_4}{L_1}\leq\frac{C_1}{C_2}\leq\frac{L_2}{L_3},\\
(\tilde{v}_1,1)\quad&\frac{C_1}{C_2}>\frac{L_2}{L_3}.
\end{aligned}
\right.
\label{policy1}
\end{equation}
If $L_1L_2<L_3L_4$, we have 
\begin{equation*}
    \frac{L_4}{L_1}>\frac{L_2+L_4}{L_1+L_3}>\frac{L_2}{L_3},
\end{equation*}
thus,
\begin{equation}
(v_1^*, v_2^*)=\left\{
\begin{aligned}
(1, \tilde{v}_2)\quad&\frac{C_1}{C_2}\leq\frac{L_2+L_4}{L_1+L_3},\\
(\tilde{v}_1,1)\quad&\text{otherwise}.
\end{aligned}
\right.
\label{policy2}
\end{equation}

\begin{figure}[htbp]
\includegraphics[width=1.0\columnwidth]{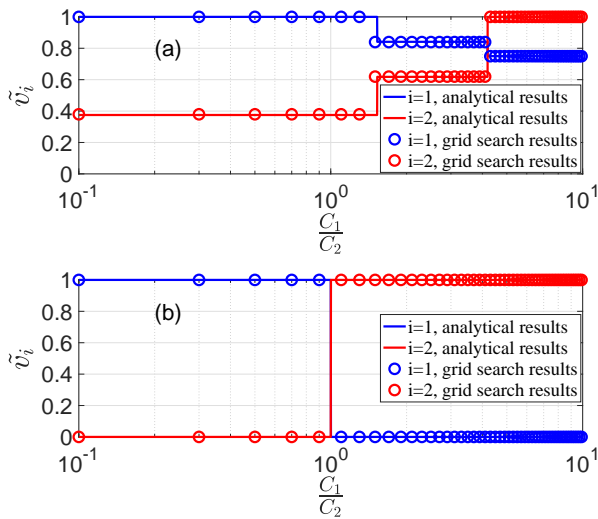}
\caption{\label{best_v}Comparison of the optimal vaccination programs derived by analytical results and grid search with respect to different vaccine cost ratios $\frac{C_1}{C_2}$ for different diseases. Here, analytical results are calculated based on Eq.~(\ref{policy1}) and Eq.~(\ref{policy2}). Parameters values (a) $R_{0,1}=10$, $R_{0,2}=8$, $\alpha_1=0.4$, $\alpha_2=0.2$; (b) $R_{0,1}=10$, $R_{0,2}=5$, $\alpha_1=0.1$, $\alpha_2=0.2$.}
\end{figure}

According to the criterion above, we can analytically derive the optimal vaccination program by substituting $(\hat{v}_1, \hat{v}_2)$ and $\tilde{v}_i$ to Eq.~(\ref{policy1}) and Eq.~(\ref{policy2}). In Fig.~\ref{best_v}, the optimal vaccination programs derived by analytical results are compared with the grid search with respect to different vaccine cost ratios $\frac{C_1}{C_2}$ for different diseases. The agreement is high. In Fig.~\ref{best_v}(a), $R_{0,1}=10$, $R_{0,2}=8$, $\alpha_1=0.4$, $\alpha_2=0.2$, so $L_1L_2=0.0607>L_3L_4=0.0221$. As a result, there are three scenarios according to the vaccine cost ratio. In Fig.~\ref{best_v}(b), $R_{0,1}=10$, $R_{0,2}=5$, $\alpha_1=0.1$, $\alpha_2=0.2$, so $L_1L_2=0.0943<L_3L_4=0.3201$. As a result, there are two scenarios according to the vaccine cost ratio.

In summary, we investigate the optimal vaccination program that can prevent the outbreaks of two SIR type diseases with cross immunity at the minimum cost. We adopt MMCA to develop an optimization framework to identify the optimal vaccination programs in various epidemiological parameter settings. We analytically derive the optimal solutions and validate the results with MC simulations and grid search. This study provides clues to design effective and efficient vaccination programs where the cross immunity between multiple diseases exists. In particular, the world is facing critical challenges in confronting the coexistence of both the novel coronavirus (COVID-19) pandemic and other major infectious diseases \cite{mateus2020selective}. With limited resources, it is important to inform the model-driven vaccination programs that can contain multiple epidemics efficiently.

%

\end{document}